\documentclass[usenatbib,fleqn]{mn2e}
\usepackage{fixltx2e,longtable,ifpdf,amsmath,amssymb}
\usepackage{times,nicefrac}
\ifpdf
  \usepackage{aeguill}
  \usepackage[pdftex]{graphicx,color}
\else
  \usepackage[dvips]{graphicx,color}
\fi

\newcommand{\teff}{\ensuremath{T_{\rm eff}}}
\newcommand{\logg}{\ensuremath{\log{g}}}
\newcommand{\mdot}{\ensuremath{\dot{M}}}
\newcommand{\msun}{\ensuremath{\mbox{M}_{\odot}}}
\newcommand{\rsun}{\ensuremath{\mbox{R}_{\odot}}}
\newcommand{\lsun}{\ensuremath{\mbox{L}_{\odot}}}

\newcommand{\rstar}{\ensuremath{R_*}}
\newcommand{\vinf}{\ensuremath{v_\infty}}

\newcommand{\vesini}{\ensuremath{v_{\rm e}\sin{i}}}
\newcommand{\veq}{\ensuremath{v_{\rm e}}}

\newcommand\zpup{\ensuremath{\zeta~\mbox{Pup}}}

\newcommand{\Ha} {\ensuremath{\mbox{H}\alpha}}

\newcommand{\kms} {\ensuremath{\mbox{km}\;\mbox{s}^{-1}}}
\newcommand{\pd}  {\ensuremath{\mbox{d}^{-1}}}
\newcommand{\pyr}  {\ensuremath{\mbox{yr}^{-1}}}

\def\fm{\hbox{$.\!\!^{\rm m}$}}

\def\utw{\smash{\rlap{\lower5pt\hbox{$\sim$}}}}
\def\udtw{\smash{\rlap{\lower6pt\hbox{$\approx$}}}}

\title[Photometry of \zpup]{Time-series photo\-metry of the O4~I(n)fp
  star $\zeta$~Puppis} \author[I.~D.~Howarth \& I.~R.~Stevens]{Ian
  D. Howarth$^1$\thanks{i.howarth@ucl.ac.uk} and
  Ian R. Stevens$^{2}$\\ $^{1}$Dept. Physics \& Astronomy, University
  College London, Gower St., London WC1E 6BT \\ $^{2}$School of
  Physics and Astronomy, University of Birmingham, Edgbaston,
  Birmingham B15 2TT}

\begin{document}


\maketitle

\label{firstpage}

\begin{abstract}
  We report a time-series analysis of the O4~I(n)fp star \zpup, based on
  optical photo\-metry obtained with the \emph{SMEI} instrument on the
  \emph{Coriolis} satellite, 2003--2006.  A single astrophysical
  signal is found, with $P = 1.780938 \pm 0.000093$~d and a mean
  semi-amplitude of $6.9 \pm 0.3$~mmag.  There is no evidence for
  persistent coherent signals with semi-amplitudes in excess of
  $\sim$2~mmag on any of the timescales previously reported in the
  literature.  In particular, there is no evidence for a signature of
  the proposed rotation period, $\sim$5.1~days; $\zeta$~Pup is
  therefore probably \emph{not} an oblique magnetic rotator.  The
  1.8-day signal varies in amplitude by a factor $\sim$2 on timescales
  of 10--100d (and probably by more on longer timescales), and
  exhibits modest excursions in phase, but there is no evidence for
  systematic changes in period over the 1000-d span of our
  observations.  Rotational modulation and stellar-wind variability
  appear to be unlikely candidates for the underlying mechanism; we
  suggest that the physical origin of the signal may be pulsation
  associated with low-$\ell$ oscillatory convection modes.
\end{abstract}


\section{Introduction}
\label{sec:intro}

There is no star in the sky that is both hotter and brighter than
\zpup\ (HD~66811; O4~I(n)fp, $V = 2.24$, \citealt{Sota11, Cousins72}).
As a result, it has long been a popular subject for the investigation
of characteristics of massive, luminous stars in general, and of their
radiatively driven stellar winds in particular, from both
observational and theoretical perspectives (e.g, \citealt{Lamers76,
  Barlow77, deLoore77, Snow80, Abbott80, Kudritzki83}; and many others
subsequently).

\subsection{Variability}

One consequence of the scrutiny under which \zpup\ has been placed
is that there are numerous reports in
the literature of low-level spectroscopic and photo\-metric variability,
across the electromagnetic spectrum.   
Although at least part of this
variability appears to be stochastic, claims of periodic or cyclical signals
can be grouped under three headings  (cf.\ the summary in
Table~\ref{tab:periods}):

\begin{table*}
\centering
\caption{Summary of periods reported for $\zeta$~Pup.  `DACs' refers to
discrete absorption components in the absorption troughs of UV P-Cygni
profiles, while `lpv' means (photo\-spheric) line-profile variability.}
\begin{tabular}{lcll}
\hline
\hline
\rule{0pt}{8pt}
Period & Epoch & \multicolumn{2}{c}{Source}\\
\hline
$5.075 \pm 0.003$~d & 1975--1976& \citet{Moffat81}    & \Ha\ absorption\\
$\sim$5.26~d        & 1986      & \citet{Balona92}    & Photo\-metry  \\
$5.21 \pm 0.71$~d   & 1995      & \citet{Howarth95}   & DACs \\
\rule{0pt}{10pt}$\sim$15~h          & 1989      & \citet{Prinja92}    & DACs \\
$19.23 \pm 0.45$~h  & 1995      & \citet{Howarth95}   & DACs \\
$19.57 \pm 0.48$~h  & 1990      & \citet{Reid96}      & \Ha\ variability \\
$16.67 \pm 0.81$~h  & 1991      & \citet{Berghofer96} & 0.1-2.4keV \\
$16.90 \pm 0.48$~h  & 1991      & \citet{Berghofer96} & \Ha\ variability \\
\rule{0pt}{10pt}$\sim$8.5~h         & 1984      & \citet{Baade86}  & lpv \\
$8.54 \pm 0.054$~h  & 1990      & \citet{Reid96}   & lpv \\
\rule{0pt}{10pt}$1.780938 \pm 0.000093$~d & 2003--2006 & This paper & Photometry \\
\hline                                        
\end{tabular}
\label{tab:periods}
\end{table*}


\medskip
\noindent(i)
\emph{8.5-hour variability:  non-radial pulsation?}  

\citet{Baade86} discovered velocity-resolved structure in the
photo\-spheric absorption lines of \zpup, with a possible 8.5-hr
periodicity in data taken in 1984/5; \citet{Reid96} found very similar
characteristics, with $P = 8.54$~hr, in spectra taken in 1990.

The observations show characteristic blue-to-red migration of bumps
and dips in the absorption-line profiles, suggesting non-radial
pulsation as the underlying physical mechanism; a tentative
identification of a sectoral mode with $\ell = -m = 2$ has been proposed
\citep{Baade88a, Reid96}.

However, while the general nature of the line-profile variability
persisted in spectra taken in 1986 and 2000, the periodic signal could
not be recovered in those  data (\citealt{Baade91}; Donati \& Howarth,
unpublished), showing it to be transient, or variable in amplitude.

\smallskip
\noindent(ii)
\emph{17--19-hr variability:  recurrent wind structures?}

Unsaturated  P-Cygni profiles of UV resonance lines in OB stars 
commonly show
`discrete absorption components' (DACs; e.g.,
\citealt{Prinja86, Kaper99}), which migrate bluewards through the
absorption troughs.  \citet{Howarth95} found a DAC recurrence timescale of
19~hours  in 16~days of IUE observations of \zpup\ taken in 1995.
Essentially the same period was recovered from 
observations of \Ha\ (a wind-formed line for \zpup)
taken in 1990 \citep{Reid96}, while   \citet{Prinja92} 
suggested a DAC recurrence timescale of around 15~hr, though
from 
only two days of
intensive IUE observations in 1989.

X-ray emission from hot stars arises in shocked material in their
stellar winds, and so is another tracer of the outflows.
\citet{Berghofer96} reported a low-amplitude 17-hr signal in 11~days'
of ROSAT data, 0.9--2~keV (undetectable at lower energies), obtained
in 1991 October.  Although this signal is not of itself particularly
persuasive (cf.\ the discussion in \citealt{Naze13}), eight days
(\emph{sic}) of contemporaneous \Ha\ spectroscopy reported by
\cite{Berghofer96} showed the same $\sim$periodic signature.  However,
\citet{Naze13} analysed a larger, XMM-Newton, dataset (16 separate
observations, 2002--2010) and found no periodic signals, concluding
that ``variations of several hours and an amplitude of a few per
cent$\ldots$is transient, at best.''

It seems plausible that all these signals may reflect a single loose, and
possibly transitory, timescale in the stellar wind.  \citet{Berghofer96} pointed
out that this timescale is ca.~2$\times$ the period found from the
absorption-line profiles, but concluded that there is no obvious
physical connection.

\smallskip
\noindent(iii)
\emph{5.1-d variability:  rotation?}

\citet{Moffat81} detected a modulation in the absorption component of the
\Ha\ P-Cygni profile in 1975--1976, consistent with a 5.1-day period. 
They interpreted this as the
stellar rotation period, suggesting that the
inner regions of the stellar wind are forced into co-rotation by a
magnetic field;  that is, that \zpup\ is an oblique magnetic
rotator. \citet{Balona92} found a marginal signal with a
semi-amplitude of $\sim$0{\fm}01 at $P \simeq 5.2$~d in Str\"{o}mgren $b$
photo\-metry  from 1986, but not from 1989;  he also noted that the dispersion in
the photo\-metry was much larger than the internal errors,
concluding that \zpup\ is an irregular microvariable.

\citet{Howarth95} reported a similar period in UV data ($P = 5.2$~d),
although this is close to the 1~\pd\ alias of the 19-hr signal found
in the same dataset;  and \citet{Baade86} made the interesting
observation that measurements of \Ha\ variability reported by
\citet{Moffat81} give a stronger signal at the mooted NRP period of 8.5~h
than the \citeauthor{Moffat81} period of 5.1~d in a
phase-dispersion-minimization periodogram (though they recognized that
the shorter period is far below the Nyquist period of the data).

\medskip

While the evidence for each of these three variability timescales is
reasonable, in every case it falls short of providing a compelling
demonstration of a persistent, coherent signal, in large part because
of the observational difficulties in obtaining extensive, well-sampled
time series with appropriate duration and cadence on a very bright
target.  A robust determination of the supposed rotation period would
be of particular value, not only because of the intrinsic interest of
testing the proposed oblique rotator model, but also because, coupled
with the observed \vesini, it would provide a strong constraint on the
stellar radius, and hence the distance, which is poorly known
(Section~\ref{sec:disco}).

\section{Time-series analysis}

\subsection{Observations}

With the foregoing in mind, we have undertaken a time-series analysis
of \zpup\ photo\-metry obtained with the Solar Mass Ejection Imager,
\emph{SMEI}.  This was one of two instruments on the
\emph{Coriolis} satellite, and incorporated three imaging cameras;
here we only use results from cameras~1 and~2, which have the best
data quality,  spanning 1077~d, grouped into four seasonal runs 
of 40, 236, 211, and 175~days
(2003
April to 2006 March), with a median cadence of 101~minutes.  
The passband was dominated by the CCD detector response, peaking at
45 per cent at 700~nm, and falling to 10 per cent at $\sim$460 and 990~nm.
Further
details on the \emph{SMEI} instrument and data-handling pipeline can
be found in \citet{Eyles03} and  \citet{Spreckley08}.

All \emph{SMEI} photo\-metry shows long-term variations of
instrumental origin (e.g., \citealt{Goss11}), which we removed with
a ten-day running-mean filter.\footnote{We performed simulations to
  verify that this has no significant impact on our sensitivity to
  $\sim$5-d signals.}  The trend-corrected observations have a
dispersion characterized by $\sigma = 19.4$~mmag; we analysed both the
full dataset, and a subset with a 3-$\sigma$ clip applied (6918 and 6855 measurements,
respectively), obtaining essentially identical results.  Numerical values
reported here are based on the clipped subset.

\begin{figure}
\centering
\includegraphics[scale=0.45,angle=0]{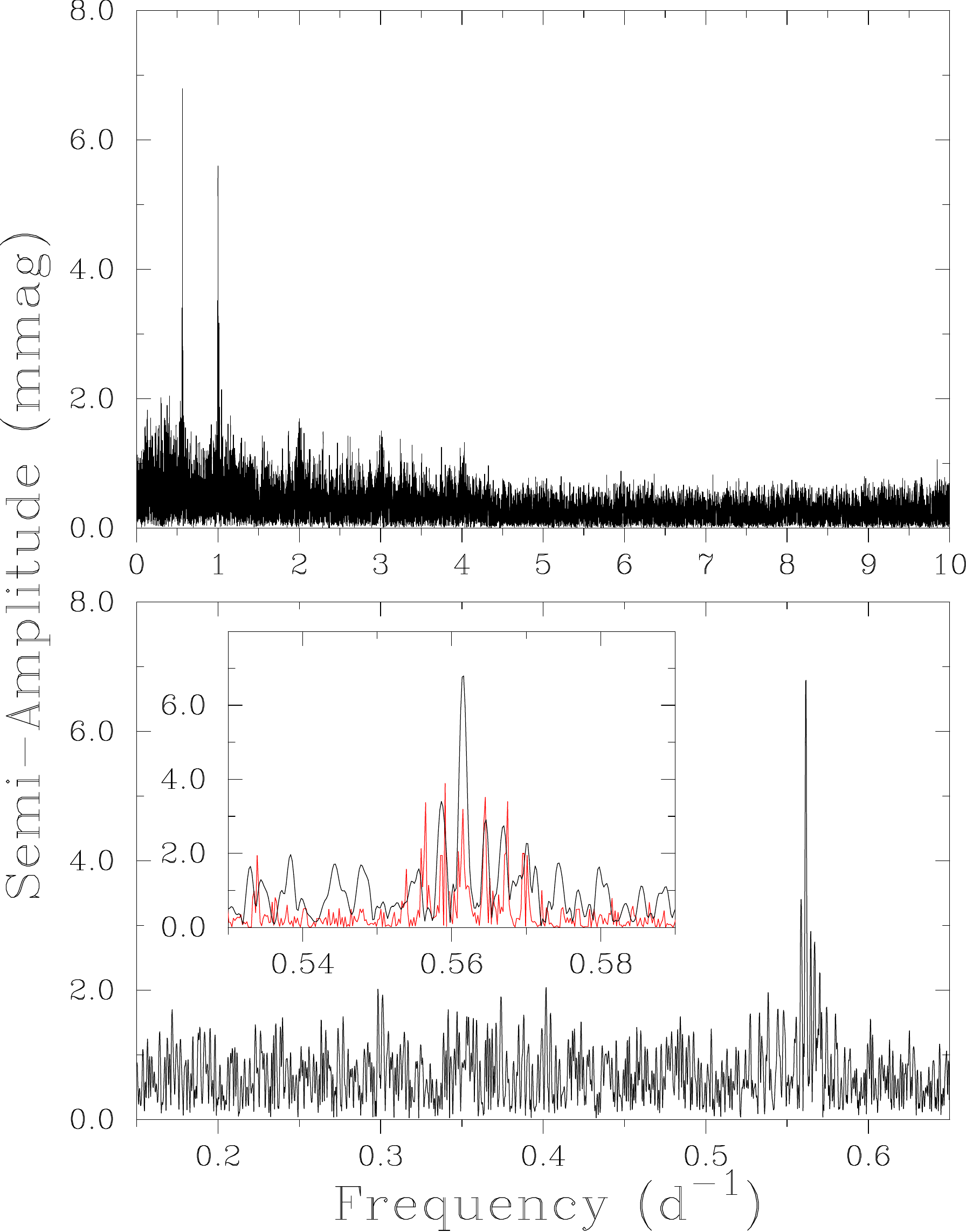}
\caption{Date-compensated discrete fourier transform of $\zeta$~Pup
  photometry.  The inset in the lower panel shows the region of the
  0.56-\pd\ signal, overlaid with a shifted version of the window
  function (in red).}
\label{fig:dcdft}
\end{figure}

\subsection{Global data properties}

\begin{figure}
\centering
\includegraphics[scale=0.35,angle=270]{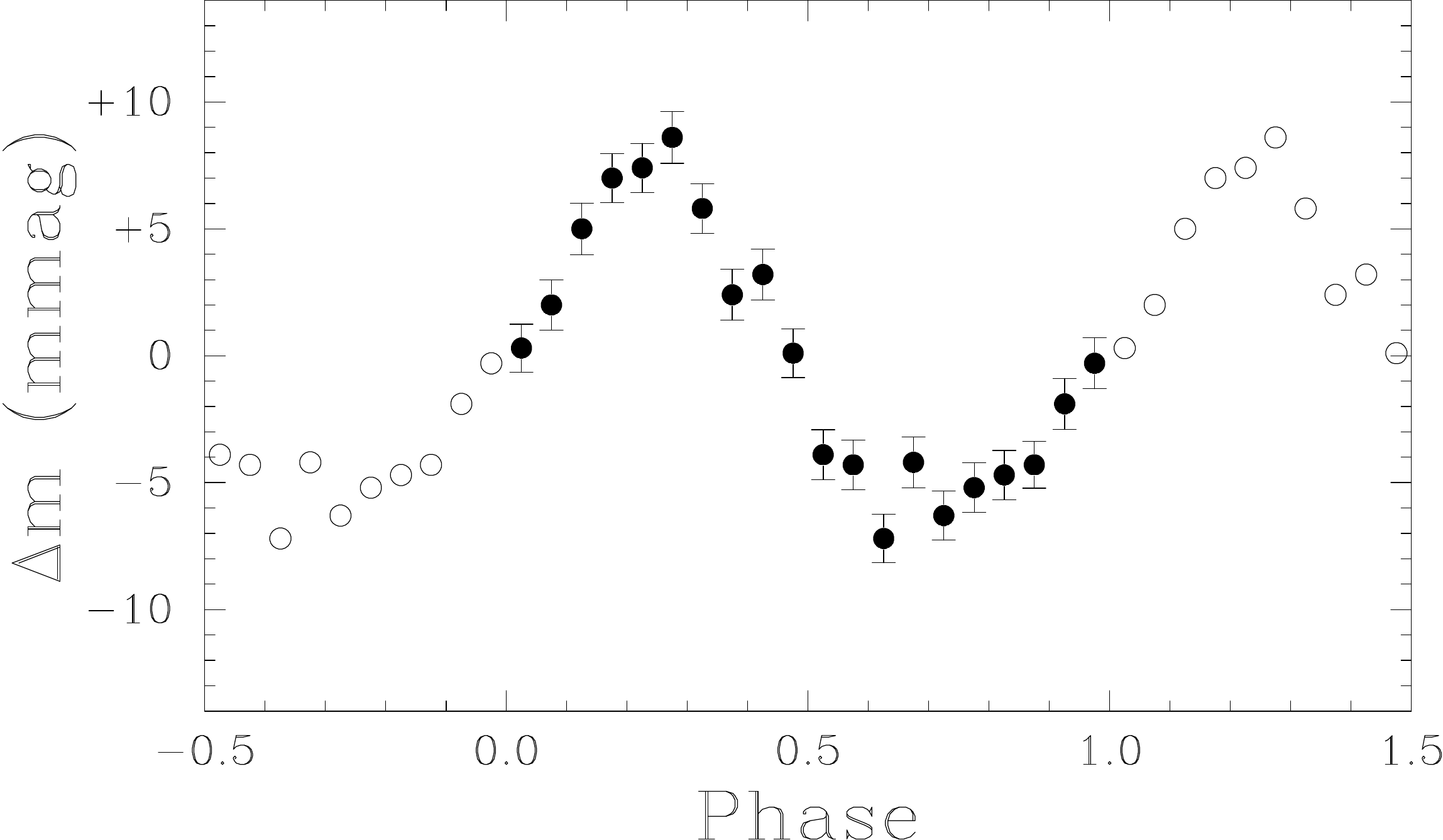}
\caption{Photo\-metry binned at $P=1.78$~d (arbitrary phasing)}
\label{fig:lcurve}
\end{figure}

Fig.~\ref{fig:dcdft} shows the date-compensated discrete component
fourier transform of the entire dataset,
over the frequency range 0--10~\pd\ (DCDFT; \citealt{Ferraz81}); the Nyquist frequency is at 7.086~\pd.
There is a single clear astro\-physical signal (in addition to an
instrumental signal at 1~\pd\ and multiples thereof), with\\ 
$\phantom{m}\nu =
0.561502\,(29)$~\pd\ $\qquad$[$P = 1.780938\,(93)$~d],\\ 
$\phantom{m}\mbox{semi-amplitude} = 6.69\,(31)$~mmag,\\ 
where bracketed values are 1-$\sigma$ uncertainties in the last
significant figures, generated by 10\,000 Monte-Carlo replications of
artificial datasets having the same input signal plus gaussian noise
(and are slightly larger than the formal single-parameter errors from a
least-squares fit of a sinusoid).    A minor periodogram peak occurs at
the first harmonic (semi-amplitude 1.6~mmag at $\nu = 1.123$~\pd);
although this would not be significant in isolation, there is
<0.1\%\ probability that a peak this strong should appear at this
particular frequency by chance.
Fig.~\ref{fig:lcurve} shows the
phased, binned data, and confirms that the signal is only slightly non-sinusoidal.

%
%
%

%

There is \emph{no} convincing signature of the mooted $\sim$5-d rotation period.
The mean semi-amplitude in the DCDFT over the frequency range
0--0.5~\pd\ is $0.66 \pm 0.37$~mmag (s.d.); the corresponding figures
over the 0.18--0.21~\pd\ range are essentially
indistinguishable ($0.66 \pm 0.36$~mmag).  There are several peaks
in the latter range with semi-amplitudes up to 1.3--1.4~mmag (which
are entirely unremarkable in the context of the broader frequency
range); the strongest, at $\nu = 0.1952$~\pd, has a semi-amplitude of
$1.4\pm 0.3$~mmag.\footnote{Of course, 
this doesn't
  represent a `3-$\sigma$ detection', because we have selected
  this frequency \emph{a posteriori} from the several thousand
  independent frequencies available.}
   We would not expect any significant change in
period, or phase, of a truly rotationally modulated signal over the
course of our observations, so our interpretation of these
results is that there is a 3-$\sigma$ upper limit of 2.3~mmag on 
the semi-amplitude of any
such signal in the period range  $P=4.8$--5.5~d.

\subsection{Transitory signals}
\label{sec:tsig}

The global DCDFT is primarily sensitive to signals at fixed phase and
period; cancellation will occur for signals which recur with different
phasing, or which drift in frequency -- circumstances that might well
be expected to apply to the $\sim$8.5-hr and $\sim$17-hr signals
discussed in Section~\ref{sec:intro}.    We therefore computed
DCDFTs for seasonal subsets of the data, and for 50-day sequences 
(starting every 25 days).   There is no suggestion of significant
power at either of the shorter periods, at any time.

\begin{figure}
\centering
\includegraphics[scale=0.50,angle=0]{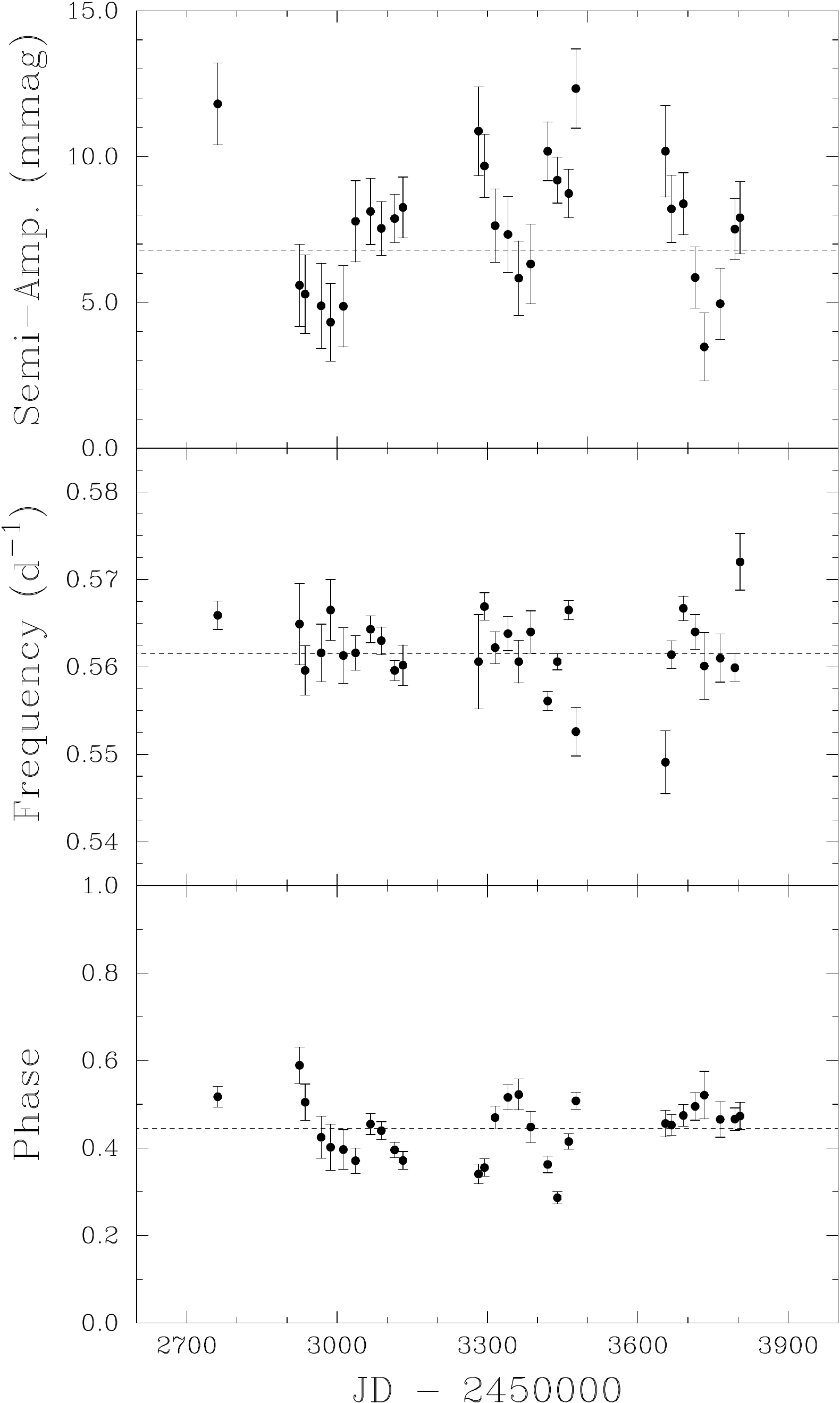}
\caption{Top two panels:  semi-amplitudes and frequencies for the 0.56~\pd\ signal from
  50-d subsets of the data, sampled every 25~d.  Mean values from the
  global analysis are shown by horizontal lines.
Bottom panel:  phase  of
  the 1.78-d signal
(referenced to mid-point of the times series), with error bars computed
  analytically (cf.~Section~\ref{sec:tsig}).}
\label{fig:sub1}
\end{figure}

The same subsets allow us to examine the coherence and stability of
the 1.78-d signal.   The semi-amplitudes and periods are summarized in
Fig.~\ref{fig:sub1} (top two panels), where the
error bars were generated analytically following \citet{Montgomery99}.
Because the points are not independent, and because the analytical
error estimates are rigorous only under restricted conditions, we
investigated the probability that the null hypotheses of constant
frequency and constant semi-amplitude can be ruled out by using a
simple Monte-Carlo approach, utilising the fact that the dispersion in
the observations is dominated by observational errors (and not by the
periodic signal).

To do this, we took the original dataset and, with observing dates fixed, shuffled
the flux values (using the Fisher-Yates algorithm;  we verified that
this removed all periodic signals).   We then planted an artificial, periodic
signal with characteristics matching those found in the original data,
and analysed the results in an identical fashion.


We find that 13\%\ of 10\,000 replications result in $\chi^2$ values as
large or larger than that actually obtained for the frequencies, but
that none of the simulations result in a $\chi^2$ value for the
amplitudes as large as that found in the data.  We conclude that this
test provides no evidence for changes in period, but that the
amplitude of the photo\-metric signal varies, by a factor $\sim$2 in our
data.


Because the semi-amplitude found for the entire dataset is consistent
with the mean of the subset semi-amplitudes, it is unlikely that there
is significant phase slippage during our observations (which would
dilute the signal in the full dataset).  We examined the coherence of
the 1.78-d signal by determining the phase, for fixed period, in the
subsets (Fig.~\ref{fig:sub1}, bottom panel).   Monte-Carlo simulations show that the
phase `wander' seen in Fig.~\ref{fig:sub1}, while of fairly low
amplitude, is too large to have arisen by chance, with $>99.9$\%\ confidence.

\section{Discussion}
\label{sec:disco}

The discovery of a strong, periodic signal in such a well-studied star
is superficially surprising, but may in part be a consequence of
$\zeta$~Pup being too bright for many programmes, and of the period
being too long to identify in short data streams.  Moreover, although
the signal was consistently present, and remained essentially
coherent, over the three years of our dataset, it isn't necessarily a
permanent feature.  Had the signal reported here been present with a
comparable amplitude in 1986--89, the time of the observations discussed by
\citeauthor{Balona92} (\citeyear{Balona92}), it would certainly have
been detected (Balona, personal communication).

Physical interpretation of the signal is handicapped by 
uncertainties in many of $\zeta$~Pup's fundamental
para\-meters, which are a direct consequence of the
uncertainty in its distance.  Although the \emph{Hipparcos} parallax yields
$d = 332 \pm 11$\,pc (\citealt{vanLeeuwen07}; see also
\citealt{Schilbach08}, \citealt{Maiz08}), this leads to estimates of
the stellar mass that are substantially smaller than generally
accepted values for O~super\-giants \citep{Bouret12}, and a case can
be made for $d \simeq 700$~pc (e.g., \citealt{Pauldrach12}).
Furthermore, \emph{if} the stellar rotation period were $P_{\rm rot} \simeq
5.1$~d, then the equatorial rotation speed\footnote{Since \zpup\ has
  one of the largest known \vesini\ values among the O~super\-giants
  (e.g., \citealt{Howarth97}), it is likely that $\sin{i} \simeq 1$.}
of $\veq \simeq 220$~\kms\ would imply $R \simeq 22\rsun$, whence $d
\simeq 540$~pc.

Consequently, while parameters that can be determined directly from
the spectrum are reasonably well established (e.g., \teff, \logg,
\vinf), the mass, radius, and luminosity are more poorly known; the
mass-loss rate has additional uncertainties arising from 
clumping in the wind.  For reference, results from two recent
analyses, obtained using independent state-of-the-art modelling tools,
are summarized in Table~\ref{tab:params}, along with ancillary
distance-dependent derived quantities.

\begin{table}
\centering
\caption{Stellar parameters, from \citeauthor{Bouret12}
  (\citeyear{Bouret12}; B12) and \citeauthor{Pauldrach12}
  (\citeyear{Pauldrach12}; P12).
$P_{\rm min}$ is the minimum rotation
period (\S\ref{sec:rot}), and $Q$ the pulsation `constant'
(\S\ref{sec:puls}).
 Bracketed values in italics are rescaled to
distances  $d = [332, 540]$~pc
(cf.~\S\ref{sec:disco});  mass-loss rates are scaled assuming
$\mdot \propto d^{3/2}$.}
\begin{tabular}{lcc}
\hline
\hline
Parameter &  B12 & P12 \\
\hline
\teff\ (kK)               & 40.0                & 40.0                \\
\logg\ (dex cgs)          & 3.64                & 3.40                \\
\vesini\ (\kms)             & 210                 & 220                \\
\vinf  (\kms)             &2300                 & 2100                \\
Adopted $d$ (pc)            & 460                 & 692               \\
\rule{0pt}{12pt}
$-\log\mdot$ (dex \msun\pyr)   & 5.70              & 4.86            \\
                          &\emph{[5.91, 5.60]} &\emph{[5.34, 5.02]}\\
$R/\rsun$                 &              18.8   & 28.0                \\
                          &\emph{[13.6, 22.1]}  & \emph{[13.4, 21.8]} \\
$\log(L/\lsun)$           & 5.91                & 6.26                \\
                          &\emph{[5.63, 6.05]}  &\emph{[5.62, 6.04]}  \\
$M/\msun$                 & 56                  & 72                  \\
                          & \emph{[29, 78]}     & \emph{[17, 44]}     \\
$P_{\rm rot}/\sin{i}$ (d)& 4.5                 &  6.4              \\
                          & \emph{[3.3, 5.3]}     & \emph{[3.1, 5.0]}     \\
$P_{\rm min}$ (d)       & 2.1                 &  3.4               \\
                          & \emph{[1.8, 2.3]}     & \emph{[2.3, 3.0]}     \\
$Q$ (d)                  & 0.16                & 0.10                 \\
                          & \emph{[0.19, 0.15]} & \emph{[0.15, 0.12]} \\
\hline                                        
\end{tabular}
\label{tab:params}
\end{table}

\subsection{Rotation}
\label{sec:rot}

For a Roche model,  the
minimum possible stellar rotation period for a positive equatorial
effective gravity is
\begin{align*}
P_{\rm min} = 3\pi\sqrt{R_{\rm eq}/g_{\rm p}},
\end{align*}
where $R_{\rm eq}$ is
  the equatorial radius and $g_{\rm p}$ is the polar gravity.   
We
  include estimates of $P_{\rm min}$ in Table~\ref{tab:params},
  taking $g_{\rm p} = g$ and $R_{\rm eq} = \sqrt{1.5}R$.

Estimates of the maximum rotation period, $P_{\rm rot}/\sin{i}$ 
(which is probably close to the true rotation period),
follow from \vesini\ and
$R$;  these are also reported in 
Table~\ref{tab:params}.

The 1.78-d photo\-metric signal is only marginally consistent 
with the shortest possible rotation period, and is 
substantially shorter 
than any plausible estimate of the true rotation period.
Eschewing numero\-logical speculation
(e.g., $P_{\rm rot} \simeq 3 P_{\rm phot}$?), this appears to rule out
rotational modulation as the cause of the photo\-metric variability.

\subsection{Wind variability}

The optical depth through
the wind can be estimated by integrating the equation of mass
continuity for an
assumed `beta' velocity law,
\begin{align*}
v(r) = v_\infty (1 - R_*/r)^\beta.
\end{align*}
The result is mildly sensitive to the adopted $\beta$ index, and to
$v_0$, the minimum velocity used for the integration, but, for $0.6
\le \beta \le 1.2$, $15 \le (v_0/\kms) \le 30$, the 
electron-scattering optical depth is
within a factor $\sqrt{3}$ of
\begin{align*}
\tau_{\rm es} \simeq
0.08 \times
\left[{    \frac{\mdot/(\msun~\pyr)}{3.5\times 10^{-6}} }\right]
\left[{ \frac{R_*/\rsun}{18.8} \,
\frac{v_\infty/(\kms)}{2200}}\right]^{-1},
\end{align*}
where we have used the mass-loss rate from \citet{Cohen10}, for their
adopted distance of 460~pc; the radius is scaled to the same distance.
(The numerical values of both \rstar\ and this $\mdot$ are directly proportional
to $d$, so their ratio is distance-independent.)

The photometric variability, if attributed to changes in continuum
optical depth of the stellar wind, would require $\Delta\tau \simeq
0.013$; that is, the wind column would have to vary by
15--20\%.  While not out of the question, such a large, periodic modulation is
unlikely to have escaped notice in previous dedicated stellar-wind
studies, and would require a driving mechanism independent of
rotation.

\subsection{Magnetic confinement}

The absence of a detectable rotational signature at the
\citet{Moffat81} 5.1-d period is noteworthy.  Their result was based
only 35 points, and they note that different amplitudes, and slightly
different best-fitting periods, pertain to two different observing
seasons, so the case for a strictly rotationally modulated signal is
not compelling, and rests largely on the near-coincidence with
estimates of the rotation period based on $\vesini$
(Table~\ref{tab:params}).  

For $\tau_{\rm es} \simeq 0.1$,  
an upper limit of $\sim$5~mmag on rotational
photometric variability implies a column-density modulation
$\lesssim$5\%\ in an asymmetric wind.  Thus if $\zeta$~Pup is indeed
an oblique magnetic rotator, then $B_{\rm p}$, the dipole field
strength at the magnetic pole, is insufficient to shape the wind
significantly.  Following \citet{udDoula02}, this implies
\begin{align*}
B_{\rm p} 
  &\lesssim 100 
\left[{    \frac{\mdot/(\msun~\pyr)}{3.5 \times 10^{-6}} 
\, \frac{\vinf/(\kms)}{2200} }\right]^{1/2}
\left[{ \frac{18.8}{R_*/\rsun} }\right] \mbox{G}.
\end{align*}
[As this paper was being prepared for submission, \citet{DavidUraz14}
  reported a 95-\%\ confidence upper limit on a dipolar field strength
  of $B_{\rm p} < 121$~G, based on one night's spectro\-polarimetric
  observations, consistent with
  our result.]

\subsection{Pulsation}
\label{sec:puls}

Pulsation would seem to be a plausible candidate mechanism for generating
the photo\-metric signal.  We include estimates of the pulsation
`constant',\footnote{Osborn's law: variables won't, constants aren't.}
\begin{align*}
Q = P \sqrt{(M/\msun)/(R/\rsun)^3}
\end{align*}
in Table~\ref{tab:params},
finding $Q \simeq 0.1$--0.2~d.

Zeta Pup is expected to be unstable to low-order (radial) $p$-mode
oscillations according to \citet{Saio11}; the luminosity:mass ratio,
$\sim4.3 \pm 0.1$ ($\log_{10}$ solar units), is large enough to suggest
the strange-mode instability associated with the iron-opacity
peak as the driving mechanism.  However, expected $Q$ values are
$\sim$0.03~d, substantially smaller than observed.

The 1.8-d period is therefore more likely to be associated with the
oscillatory convection modes discussed by \citet{Saio11}.
The $Q$ values for low-order ($\ell = 1, 2$) modes, which are expected
to be the most
readily observable, are $\sim$0.2--0.3 for the models most likely to be
relevant to \zpup\footnote{Saio (personal communication) points out that $Q \simeq 0.06$--0.08 for
  Geneva models with initial masses $\sim$40--50\msun\ during core
  helium burning, when they return to the vicinity of the main
  sequence following an excursion to the red in the
  Hertzsprung-Russell diagram.  In principle, this could be consistent
  our results, particularly since the models predict masses at this
  stage that are $\sim$half the zero-age main-sequence values, with a
  commensurate reduction in the `observed' $Q$.  However, although it
  is generally accepted that CNO-processed material is exposed at the
  surface of $\zeta$~Pup (e.g., \citealt{Bouret12}), surface
  abundances have not progressed to the strongly non-solar values
  predicted at this stage in evolutionary models by, e.g.,
  \citet{Ekstrom12}.  Moreover, single-star evolutionary models show
  considerable rotational spindown over the main-sequence phase;  the
  exceptionally high \vesini\ observed for \zpup\ therefore argues for
  it being core hydrogen burning.  Merger models offer an alternative
  mechanism for generating rapid rotation, but core hydrogen burning
  appears to be in effect even for the merger model discussed by
  \citet{Pauldrach12}.}
(core hydrogen burning, solar metallicity, $M_{\rm
  ZAMS}\simeq 60$~\msun),
 reasonably consistent with our observed
value.

We arrive at this conclusion in part through the application of
Holmes' maxim (\citealt{Doyle92}, p.~524), as the match in $Q$ is far
from perfect, and the inclusion of rotation in the models is liable to
shift the predicted frequencies to larger values.  Furthermore, in
displaying a single, strong signal, the frequency spectrum for
\zpup\ differs from those found for other early-type O stars, which
appear to have power spectra dominated by red noise \citep{Blomme11},
although the available sample is small. We speculate that a range of
modes may actually be excited in \zpup, and that we have seen just the `tip of
the iceberg'.

\section{Summary}

Four years' of $\zeta$~Pup photometry from the \emph{SMEI} instrument,
2003--6, reveals a single astrophysical signal, with $P = 1.780938 \pm
0.000093$~d and a mean semi-amplitude of $6.9 \pm 0.3$~mmag.  The
period appears too short to be rotational, and the amplitude too large
to arise through wind variability.  We therefore tentatively attribute
the signal to pulsation, possibly associated with low-order
oscillatory convection modes.  Any signal associated with a mooted
\mbox{$\sim$5-d} rotation period had a semi-amplitude $<2.3$~mmag at
the time of our observations, with 3-$\sigma$ confidence.

\section*{Acknowledgements}

We thank Hideyuki Saio for suggestions, Vino Sangaralingam for
assistance with initial data processing, and our referee, Luis Balona,
for helpful remarks.

\bibliographystyle{mn2e}

\bibliography{IDH}

\appendix

\label{lastpage}

\end{document}